\documentclass[10pt,english]{article}
\usepackage{times}
\usepackage[T1]{fontenc}
\usepackage[latin1]{inputenc}
\usepackage{amsmath}
\usepackage{amssymb}

\makeatletter

\newcommand{\noun}[1]{\textsc{#1}}
\providecommand{\tabularnewline}{\\}

\usepackage{slashed}

\usepackage{babel}
\makeatother
\begin{document}

\title{\textbf{CHARGED AND NEUTRAL CURRENTS IN A 3-3-1 MODEL WITH RIGHT-HANDED
NEUTRINOS}}

\author{\noun{ADRIAN} PALCU}

\date{\emph{Department of Theoretical and Computational Physics - West
University of Timi\c{s}oara, V. P\^{a}rvan Ave. 4, RO - 300223 Romania}}

\maketitle
\begin{abstract}
The charged and the neutral currents are obtained by using a formal
algebraical approach (developed and applied by the author) within
the exact solution of a 3-3-1 gauge model with right-handed neutrinos.
The entire Standard Model phenomenology is recovered without imposing
any supplemental condition, but only by choosing an adecquate set
of parameters from the very beginning of the calculus. A new and rich
phenomenology regarding the particles and their currents occurs as
well. The appealing feature of our results resides in the exact expressions
of the currents which need not the adjustment usually due to the small
mixing angle $\phi$ between neutral bosons $Z$ and $Z^{\prime}$
(like in the most of the papers in the literature treating the same
issue). The required mixing was considered and aleready performed
as an intermediate step by the solving method itself, since the physical
eigenstates of those bosons were determined and then identified in
the neutral currents. 


\end{abstract}

\section{Introduction}

Among the suitable extentions of the Standard Model (SM), the so called
''3-3-1 models'' have gained a wide reputation in the last decade.
An impressive amount of research papers has been devoted to this issue,
both from theoretical and experimental points of view. They try to
better explain the fundamental electroweak interaction of particles
within a larger framework than that of the SM, while keeping unaltered
the whole phenomenology of the strong interaction as it was established
by the Quantum Chromodynamics (QCD)\emph{.} The color group of QCD
remains an exact $SU(3)_{c}$ symmetry with its known features (quark
families, massless gluons, 3 color degree of freedom for each quark,
the running coupling of the theory, ''asympthotic freedom'' and
''infrared slavery'' of quarks, etc). The gauge group of the new
theory is now $SU(3)_{c}\otimes SU(3)_{L}\otimes U(1)_{X}$. Within
the electro-weak sector it undergoes a spontaneous symmetry breakdown
(SSB) via an adecquate Higgs mechanism. Therefore, the one finally
remains with the residual symmetry $SU(3)_{c}\otimes U(1)_{em}$. 

The particle content of the first version of such 3-3-1 models proposed
by Felice Pisano, Vicente Pleitez and Paul Frampton \cite{key-1,key-2}
included quarks with exotic electric charges, namely $+5e/3$ and
$-4e/3$, and developed a rich phenomenology at not very high energy
scale. Shortly afterwards, a new and interesting version of such 3-3-1
models - without exotic electric charges, this time - emerged \cite{key-3}
in the literature. It has been developed ever since by papers of Hoang
Ngoc Long and his collaborators \cite{key-4} - \cite{key-13} by
incorporating in the theory the right-handed neutrinos as the second
neutral component of each lepton triplet in the model. Meanwhile,
a rigourous and exhaustive classification of the models that cancel
the axial anomalies by an interplay between fermion families has been
made \cite{key-14} and a special atention has been paid to various
Higgs sectors \cite{key-15} - \cite{key-24} that can supply the
appropriate SSB in two steps $331\rightarrow321\rightarrow31$. Some
issues like the CP-phases violation \cite{key-25} - \cite{key-31},
the flavor changing neutral currents (FCNC) \cite{key-3,key-25},
\cite{key-31} - \cite{key-36}, or the fermion mass generation mechanisms
\cite{key-12}, \cite{key-37} - \cite{key-46} have been addressed
dealing with different representations for lepton and quark families. 

Remarkable successes of this class of gauge models include the prediction
for the number of fermion generations (restricted to 3 families only)
required by the renormalization criterion fulfilled by the anomalies
cancellation and the explicit quantization in the electromagnetic
sector \cite{key-47} - \cite{key-54}. 

One of the most stringent features that any extention of the SM has
to enclose is that of the mass generation and mixing pattern for the
three flavor neutrinos, since the neutrino oscillation phenomenon
was experimentally confirmed. Therefore, many attempts to solve this
issue have been proposed succesively. Assuming that neutrinos develop
tiny masses, these attempts followed three possible lines: (i) the
tree level neutrino masses \cite{key-55,key-56} with or without enforcing
additional global symmetries ($A_{4}$ or $L_{e}-L_{\mu}-L_{\tau}$)
on the resulting mass matrix \cite{key-57,key-58}; (ii) radiative
mechanisms \cite{key-59} - \cite{key-65}, \cite{key-13} championed
mainly by the papers of Teruyuki Kitabayashi; and (iii) various see-saw
approaches \cite{key-66} - \cite{key-69}. 

At a certain moment, this great variety of ways to deal with 3-3-1
models called for a systematic approach, namely an appropriate general
method to solve any gauge model with SSB mechanism in order to obtain
its neutral and charged currents, the boson, quark and lepton mass
spectrum and the anomaly-free set of fermion representations. This
goal is achieved in a very elegant algebraic manner with the paper
of Ion Cot\u{a}escu \cite{key-70} providing the modern elementary
physics with an efficient vehicle that allows for any gauge model
$SU(3)_{c}\otimes SU(N)_{L}\otimes U(1)_{X}$ to arrive to a complete
and exact description of its phenomenology just by choosing a set
of parameters at the beginning of the calculus. This way was followed
by the author (Refs. \cite{key-56}, \cite{key-58}, \cite{key-65}
and \cite{key-69}) in deriving certain features - mostly regarding
the neutrino phenomenology - of a particular class of 3-3-1 models
that includes right-handed neutrinos (initially considered in Refs.
\cite{key-3} - \cite{key-5}).

In this paper, we revisit the 3-3-1 model with right-handed neutrinos
within the formalism of the general method \cite{key-70} of exactly
solving models with high symmetries. We intend to make small changes
and adjustments in our results (\cite{key-56}) in order to prove
that there is no need for any supplemental condition once the set
of parameters was choosen at the beginning. This choice enforces a
certain representation content in the fermion sector and recovers
naturally the whole SM phenomenology after the SSB up to the residual
universal symmetry $U(1)_{em}$ took place. In Ref. \cite{key-56}
we gave the resulting charges of only the leptons with respect to
the neutral bosons $Z$ and $Z^{\prime}$(Tables 1 and 2 therein),
while the coupling coefficients of the quarks with respect to the
same bosons were completely ignored. They could have come out different
compared to those proposed in the previous papers \cite{key-3,key-5}
and thus could have developed a quite different phenomenology. We
prove in the following that this does not happen, namely the exact
charged and neutral currents obtained by our method meet the established
ones. Note that our results need no the adjustment due to the mixing
angle $\phi$ between neutral bosons $Z$ and $Z^{\prime}$, like
in the most previous papers. This feature comes from the fact that
the method itself includes this mixing as a compulsory step in determining
the physical (mass) eigenstates for the above-mentioned neutral bosons.
Moreover, our conclusion is that the general method mandatorily implies
to recover this way the SM content for the ''traditional'' leptons
and quarks. At the same time, it meets the well-known relation between
couplings ($g$ and $g^{\prime}$) on algebraical grounds not by matching
them in the first step of the breaking symmetry (as in other papers,
for example Refs. \cite{key-5}). 

The paper is organized as follows. Section 2 is devoted to the general
method of solving gauge models with high symmetries which is then
applied in Section 3 to a 3-3-1 gauge model by properly choosing the
set of parameters. Here are also derived the main results of the paper,
namely the exact expressions of the charged and neutral currents in
the model of interest. The final section sketches the main phenomenological
consequences and comments on their implications for the particle physics,
validating the theoretical method employed here.

\section{The General Method of Solving Gauge Models}

We recall in this section the main results of the exactly solving
method for generalized $SU(n)_{L}\otimes U(1)_{Y}$ electro-weak gauge
models, in the elegant and efficient algebraical manner shown in Ref.
\cite{key-70}. 

First of all, the unitary irreducible representations (irreps) with
respect to this gaugeable symmery are constructed taking into account
that the fermion multiplets of the theory obey the two fundamental
representations of the $SU(n)_{L}$: $\mathbf{n}$ and $\mathbf{n^{*}}$.
Let the coupling constants of the two groups be: $g$ for $SU(n)_{L}$,
and $g^{\prime}$ for the $U(1)_{Y}$ with $\xi=\xi^{+}\in su(n)$
the element of the algebra and $y_{ch}$ the character of the chiral
$U(1)_{Y}$ group parametrized by $\xi^{0}$. Under these conditions,
the fermion multiplets transform like

\begin{equation}
\psi\rightarrow e^{-i(g\xi+g^{\prime}y_{ch}\xi^{0})}\psi\label{Eq. 1}\end{equation}
For the simplicity of the calculus the general method deals with $y=y_{ch}g^{\prime}/g$. 

The scalar sector of any ''pure left'' gauge model must consist
of $n$ Higgs multiplets $\phi^{(1)}$, $\phi^{(2)}$, ... , $\phi^{(n)}$
satisfying the orthogonal condition $\phi^{(i)+}\phi^{(j)}=\phi^{2}\delta_{ij}$
in order to eliminate unwanted Goldstone bosons that could survive
the SSB. Here $\phi$ is a gauge-invariant real field variable and
$n$ is the dimension of the fundamental irreducible representation
of the gauge group. The parameter matrix $\eta=\left(\eta_{0},\eta{}^{(1)},\eta{}^{(2)},...,\eta{}^{(n)}\right)$
with the property $Tr\eta^{2}=1-\eta_{0}^{2}$ has to be introduced
in order to obtain a VEV splitting and consequently a non-degenerate
boson mass spectrum. Then, the Higgs Lagrangian density (Ld) stands:

\begin{equation}
\mathcal{L}_{H}=\frac{1}{2}\eta_{0}^{2}\partial_{\mu}\phi\partial^{\mu}\phi+\frac{1}{2}\sum_{i=1}^{n}\left(\eta{}^{(i)}\right)^{2}\left(D_{\mu}\phi^{(i)}\right)^{+}\left(D^{\mu}\phi^{(i)}\right)-V(\phi)\label{Eq. 2}\end{equation}
where $D_{\mu}\phi^{(i)}=\partial_{\mu}\phi^{(i)}-ig(A_{\mu}+y^{(i)}A_{\mu}^{0})\phi^{(i)}$
act as covariant derivatives of the model. More details regarding
the construction of the fermion sector, boson sector and scalar sector
respectively, can be found in the Ref. \cite{key-70}

After the SSB which straightforwardly leads to the residual universal
symmetry $U(1)_{em}$, the boson mass matrices take the forms: \begin{equation}
M_{i}^{j}=\frac{1}{2}g\left\langle \phi\right\rangle \sqrt{\left[\left(\eta^{(i)}\right)^{2}+\left(\eta^{(j)}\right)^{2}\right]}\label{Eq. 3}\end{equation}
for the non-diagonal gauge bosons (which usually are charged but,
in the 3-3-1 model under consideration here, one of them will be also
neutral) and:

\begin{equation}
\left(M^{2}\right)_{ij}=\left\langle \phi\right\rangle ^{2}Tr\left(B_{i}B_{j}\right)\label{Eq. 4}\end{equation}
with: \begin{equation}
B_{i}=g\left[D_{i}+\nu_{i}\left(D\nu\right)\frac{1-\cos\theta}{\cos\theta}\right]\eta\label{Eq. 5}\end{equation}
for the diagonal gauge bosons of the model. The last ones are neutral
without exception. The $\eta$ parameter diagonal matrix coming from
the scalar sector essentially determines the mass spectrum of the
model, while $\theta$ is the rotation angle around the versor $\nu$
orthogonal to the electromagnetic direction in the parameter space
(for a detailed analysis the reader is also reffered to Ref. \cite{key-70}).
The versor condition holds $\nu_{i}\nu^{i}=1.$ For the concrete model
we will work on, $D$s are the Hermitian diagonal generators (determining
the Cartan subalgebra) of the $SU(3)_{L}$ group, \emph{i.e.} $D_{1}=T_{3}$
and $D_{2}=T_{8}$ connected to the Gell-Mann matrices in the manner
$T_{a}=\lambda_{a}/2$. That is $D_{1}=(1/2)diag(1,-1,0)$ and $D_{2}=(1/2\sqrt{3})diag(1,1,-2)$.

After we extracted the electromagnetic potential (the field $A_{\mu}^{em}$
that is massless), a special $SO(n-1)$ transfomation $\omega$ remains
to be determined in each particular case, in order to bring the mass
matrix into the physical basis $(A_{\mu}^{em},Z_{\mu},Z_{\mu}^{\prime})$
so that the masses of physical neutral bosons are just the eigenvalues
of the diagonal form of the matrix (4). The generalized Weinberg transformation
(gWt) designed to do this job can be defined as:

\begin{equation}
A_{\mu}^{0}=A_{\mu}^{em}\cos\theta-\nu_{i}\omega_{\cdot j}^{i\cdot}Z_{\mu}^{j}\sin\theta\label{Eq. 6}\end{equation}

\begin{equation}
A_{\mu}^{k}=\nu^{k}A_{\mu}^{em}\sin\theta+\left[\delta_{i}^{k}-\nu^{k}\nu_{i}\left(1-\cos\theta\right)\right]\omega_{\cdot j}^{i\cdot}Z_{\mu}^{j}\label{Eq. 7}\end{equation}

Furthermore, the charges of the particles can be identified as coupling
coefficients of the currents with respect to the physical bosons (considered
in their mass eigenstates basis). In a certain representation $\rho$
the exact expressions of the charge operators have the following diagonal
forms \cite{key-70}:

\begin{equation}
Q^{\rho}(A_{\mu}^{em})=g\left[(D^{\rho}\nu)\sin\theta+y_{\rho}\cos\theta\right]\label{Eq. 8}\end{equation}

\begin{equation}
Q^{\rho}(Z_{\mu}^{i})=g\left[D_{k}^{\rho}-\nu_{k}(D^{\rho}\nu)(1-\cos\theta)-y_{\rho}\nu_{k}\sin\theta\right]\omega_{\cdot i}^{k\cdot}\label{Eq. 9}\end{equation}

All the charges of the particles in a certain multiplet can be obtained
straightforwardly from this point just by taking into consideration
the multiplet's own representation $\rho$ with its hypercharge $y_{\rho}$value
(keeping in mind that at the end of the calculus one has to replace
$y_{\rho}$ with $y_{ch}^{\rho}(g^{\prime}/g)$in order to deal with
the veritable chiral character) and by making an appropriate choice
for the versor set $\nu_{i}$. Note that the parameter set of the
Higgs sector is not involved in the expressions of the neutral currents,
except for the $\theta$ angle, but this is very precisely connected
to the Weinberg angle $\theta_{W}$since it determines the desired
electric charges of the multiplets.

\section{The Exact Solution of the 3-3-1 Model with Right-Handed Neutrinos
Revisited}

\subsection{Parameters}

The general method of exactly solving chiral gauge models with high
symmetries calls for a suitable set of parameters, choosen at the
very beginning of the procedure. Since the Higgs sector of any 3-3-1
model consists of three Higgs triplets $\phi^{(1)}$, $\phi^{(2)}$
and $\phi^{(3)}$which ensure a good non-degenerate boson mass spectrum,
one choses $\eta^{2}=(1-\eta_{0}^{2})diag\left[(a+b)/2,1-a,(a-b)/2\right]$
where, for the moment, $a$ and $b$ are arbitrary non-vanishing real
parameters. Obviously, $\eta_{0},a\in[0,1)$. 

Regarding the gWt where the parameter set $(g,\theta,\nu)$ is manifest,
the following option seems quite natural: $\nu_{0}=0$, $\nu_{1}=0$,
and $\nu_{2}=1$ (not $\nu_{2}=-1$, as it was chosen in Ref. \cite{key-56}).
It is also reasonable to consider that the coupling constant $g$
of the 3-3-1 model coincides with the first coupling constant of the
SM. 

Under these circumstances, the seven initial free parameters of the
exact solution of the 3-3-1 model with right-handed neutrinos are:

\begin{equation}
(a,b,g_{SM},\theta,\nu_{0}=0,\nu_{1}=0,\nu_{2}=1)\label{Eq.10}\end{equation}

The relation between the $\theta$ angle from our parametrization
and $\theta_{W}$ from SM naturally arises from the constraint $e=g\sin\theta_{W}$,
since the fundamental triplet of the theory (eq. (8) with $y_{\rho}=0$)
must denote a quark generation. As it is expected $\sin\theta=\frac{2}{\sqrt{3}}$$\sin\theta_{W}$
holds. Using this particular value of $\theta$ in diagonalization
of the neutral boson mass matrix (4) one remains with only one parameter
out of the two initial ones, since the SM condition between the masses
of the charged and neutral bosons is fulfiled in the manner $m_{Z}^{2}=m_{W}^{2}/\cos^{2}\theta_{W}$. 

Finally, one deals with a very appealing set of parameters:

\begin{equation}
(e,\theta_{W},a,\nu_{2}=1)\label{Eq.11}\end{equation}
It has been reduced from the initial number of seven to only one parameter,
namely $a$, since the first two ones in (11) come from SM and they
are already well established.

\subsection{Boson Mass Spectrum}

We review the mass spectrum of the gauge bosons as it results from
Ref. \cite{key-56} without insisting on the computing details (explicitly
developed therein). We consider throughout the paper: $m^{2}=g^{2}\left\langle \phi\right\rangle ^{2}(1-\eta_{0}^{2})/4$.
Then, the masses of the gauge bosons are: $m_{W}^{2}=m^{2}a$, $m_{X}^{2}=m^{2}\left(1-a/2\cos^{2}\theta_{W}\right)$
and $m_{Y}^{2}=m^{2}\left[1-a(1-\tan^{2}\theta_{W})/2\right]$ for
the charged ones, while $m_{Z}^{2}=m^{2}a/\cos^{2}\theta_{W}$ is
the mass of the Weinberg boson from SM and $m_{Z^{\prime}}^{2}=m^{2}\left[4\cos^{2}\theta_{W}-a\left(3-4\sin^{2}\theta_{W}+\tan^{2}\theta_{W}\right)\right]/\left(3-4\sin^{2}\theta_{W}\right)$
stands for the mass of the new neutral boson, specific to this 3-3-1
model.

The mass scale is now just a matter of tuning the parameter $a$ in
accordance with the possible values for $\left\langle \phi\right\rangle $.
This issue could be elucidated only when we will have a precise experimental
measurement of the new bosons predicted by the 3-3-1 theory. Certain
constraints could be supplied, for instance, by the neutrino phenomenology.
This kind of approach was proposed with notable results in the previous
papers of the author (see, for example, Refs. \cite{key-56,key-58,key-69})

\subsection{The General Weinberg Transformation}

In our 3-3-1 model, assuming the above versor choice, the gWt reads

\[
A_{\mu}^{0}=A_{\mu}^{em}\cos\theta-\left(\omega_{\cdot1}^{2\cdot}Z_{\mu}^{1}+\omega_{\cdot2}^{2\cdot}Z_{\mu}^{2}\right)\sin\theta\]

\begin{equation}
A_{\mu}^{3}=\omega_{\cdot1}^{1\cdot}Z_{\mu}^{1}-\omega_{\cdot2}^{1\cdot}Z_{\mu}^{2}\label{Eq.12}\end{equation}

\[
A_{\mu}^{8}=A_{\mu}^{em}\sin\theta+\left(\omega_{\cdot1}^{2\cdot}Z_{\mu}^{1}+\omega_{\cdot2}^{2\cdot}Z_{\mu}^{2}\right)\cos\theta\]
The transformation $\omega$ reduces here to a simple rotation (of
angle $\phi$, in order to keep the notation existing in the literature
for this mixing). Bearing in mind that the neutral boson mass matrix
to be diagonalized has become

\begin{equation}
M^{2}=m^{2}\left(\begin{array}{ccc}
\left[1-\frac{a}{2}\left(\frac{1-2s^{2}}{1-s^{2}}\right)\right] &  & -\frac{1}{\sqrt{3-4s^{2}}}\left[1-\frac{a}{2}\left(\frac{3-2s^{2}}{1-s^{2}}\right)\right]\\
\\-\frac{1}{\sqrt{3-4s^{2}}}\left[1-\frac{a}{2}\left(\frac{3-2s^{2}}{1-s^{2}}\right)\right] &  & \frac{1}{3-4s^{2}}\left[1+\frac{3}{2}a\left(\frac{1-2s^{2}}{1-s^{2}}\right)\right]\end{array}\right)\label{Eq.13}\end{equation}
one obtains for the specific rotation $\omega$ the expression

\begin{equation}
\omega=\frac{1}{2\sqrt{1-\sin^{2}\theta_{W}}}\left(\begin{array}{ccc}
\sqrt{3-4\sin^{2}\theta_{W}} &  & -1\\
\\1 &  & \sqrt{3-4\sin^{2}\theta_{W}}\end{array}\right)\label{Eq.14}\end{equation}
which will lead - when inserted in gWt (12) - to the correct charges
of the fermions, and subsequently to the expected neutral charges
(the couplings of the neutral currents) for all the particles in the
theory.

\subsection{Currents}

We observe that the fundamental multiplet (with $y_{\rho}=0$), must
obey the fundamental $\mathbf{3^{*}}$representation in order to denote
a quark multplet. It is characterized by the generators $D_{i}^{\mathbf{3^{*}}}=-D_{i}^{\mathbf{3}}$,
that means the two Hermitian diagonal generators stand as $T_{3}^{\mathbf{3^{*}}}=-(T_{3}^{\mathbf{3}})^{T}$
and $T_{8}^{\mathbf{3^{*}}}=-(T_{8}^{\mathbf{3}})^{T}$ respectively.

\subsubsection{Electric charges}

Equation (8) - via the particular relation between parameters $e,g,\theta_{W}$
known from SM - supplies the concrete shape of the electric charge
operator. This is:

\begin{equation}
Q^{\rho}(A_{\mu}^{em})=\frac{2e}{\sqrt{3}}\left[T_{8}^{\rho}+y_{ch}^{\rho}\left(\frac{g^{\prime}}{g}\right)\frac{\sqrt{3-4\sin^{2}\theta_{W}}}{\sin\theta_{W}}\right]\label{Eq.15}\end{equation}

From the fundamental representation $\rho=(\mathbf{3^{*}},0)$ one
can get straightforwardly the electric charge operator in particular
irreps $\rho=(\mathbf{3},+\frac{1}{3})$ and $\rho=(\mathbf{3},-\frac{1}{3})$.
Therefore, after some algebra, we have:

\begin{equation}
Q^{(\mathbf{3^{*}},0)}(A_{\mu}^{em})=Diag\left(-\frac{e}{3},-\frac{e}{3},+\frac{2e}{3}\right)\label{Eq.16}\end{equation}

\begin{equation}
Q^{(\mathbf{3},+\frac{1}{3})}(A_{\mu}^{em})=Diag\left(+\frac{2e}{3},+\frac{2e}{3},-\frac{e}{3}\right)\label{Eq.17}\end{equation}

\begin{equation}
Q^{(\mathbf{3},-\frac{1}{3})}(A_{\mu}^{em})=Diag\left(0,0,-e\right)\label{Eq.18}\end{equation}
Equations (16), (17) and (18) recover well-known values for the electric
charges of elementary particles. They are fulfiled if and only if 

\begin{equation}
\frac{g^{\prime}}{g}=\frac{\sqrt{3}\sin\theta_{W}}{\mathbf{\sqrt{3-4\sin^{2}\theta_{W}}}}\label{Eq.19}\end{equation}
holds. We obtained the formula connecting the two couplings of the
theory (also present in previous papers devoted to 3-3-1 models with
right-handed neutrinos), as a result of enforcing certain restrictions
on the eigenvalues of the charge operator and not - as usual - by
couplings match at the first step of SSB. Actually, our method avoids
the steps of the SSB, since it reaches the universal residual symmetry
$U(1)_{em}$ at once, by parametrizing the vector space of the scalar
fields. 

Before proceeding to determine the neutral charges for the above considered
irreps, we can identify the fermion families to be described by such
irreps. They are:

\paragraph{Lepton families}

\begin{equation}
\begin{array}{ccccc}
f_{\alpha L}=\left(\begin{array}{c}
\nu_{\alpha}^{c}\\
\nu_{\alpha}\\
e_{\alpha}\end{array}\right)_{L}\sim(\mathbf{1,3},-1/3) &  &  &  & \left(e_{\alpha L}\right)^{c}\sim(\mathbf{1},\mathbf{1},-1)\end{array}\label{Eq.20}\end{equation}

\paragraph{Quark families}

\begin{equation}
\begin{array}{ccc}
Q_{iL}=\left(\begin{array}{c}
D_{i}\\
-d_{i}\\
u_{i}\end{array}\right)_{L}\sim(\mathbf{3,3^{*}},0) &  & Q_{3L}=\left(\begin{array}{c}
T\\
t\\
b\end{array}\right)_{L}\sim(\mathbf{3},\mathbf{3},-1/3)\end{array}\label{Eq.21}\end{equation}

\begin{equation}
\begin{array}{ccc}
(b_{L})^{c},(d_{iL})^{c}\sim(\mathbf{3},\mathbf{1},-1/3) &  & (t_{L})^{c},(u_{iL})^{c}\sim(\mathbf{3},\mathbf{1},+2/3)\end{array}\label{Eq.22}\end{equation}

\begin{equation}
\begin{array}{ccccccccc}
(T_{L})^{c}\sim(\mathbf{3,1},+2/3) &  &  &  &  &  &  &  & (D_{iL})^{c}\sim(\mathbf{3,1},-1/3)\end{array}\label{Eq.23}\end{equation}
with $i=1,2$. 

In the representations presented above we assumed, like in majority
of the papers in the literature, that the third generation of quarks
transforms differently from the other two ones. This could explain
the unusual heavy masses of the third generation of quarks. Tha capital
letters denote the exotic quarks included in each family.

With this assignment the fermion families cancel all the axial anomalies
by just an interplay between them, although each family remains anomalous
by itself.

\subsubsection{Neutral charges}

It is obvious that one of the two neutral bosons ($Z_{\mu}^{1}$and
$Z_{\mu}^{2}$) of the present model is nothing but the neutral boson
from SM. Since its mass was calculated in this respect, we expect
that at least its couplings with the ''classical'' particles be
the same with those in the SM. Which one of the two neutral bosons
fulfils this requirement? The answer is given by the values of the
neutral charges for the leptons. It would be natural the following
outcome: $Q(Z_{\mu}^{0})=\frac{e}{\sin2\theta_{W}}diag(unkown,1,2\sin^{2}\theta_{W}-1)$
for the lepton family $f_{\alpha L}\sim(\mathbf{1,3},-1/3)$ in Eq.
(20) and $Q(Z_{\mu}^{0})=\frac{e}{\sin2\theta_{W}}2\sin^{2}\theta_{W}$
for the right-handed charged lepton $\left(e_{\alpha L}\right)^{c}\sim(\mathbf{1},\mathbf{1},-1)$
in order to keep consistency with SM phenomenology \cite{key-71}.

The neutral charges corresponding to the neutral bosons $Z_{\mu}^{1}$and
$Z_{\mu}^{2}$ in the representation $\rho$ are expressed by:

\begin{equation}
Q^{\rho}(Z_{\mu}^{1})=\frac{e\sqrt{3-4\sin^{2}\theta_{W}}}{\sin2\theta_{W}}\left(T_{3}^{\rho}+\frac{1}{\sqrt{3}}T_{8}^{\rho}-y_{ch}^{\rho}\frac{2\sin^{2}\theta_{W}}{3-4\sin^{2}\theta_{W}}\right)\label{Eq.24}\end{equation}

\begin{equation}
Q^{\rho}(Z_{\mu}^{2})=\frac{e}{\sin2\theta_{W}}\left(-T_{3}^{\rho}+\frac{3-4\sin^{2}\theta_{W}}{\sqrt{3}}T_{8}^{\rho}-y_{ch}^{\rho}2\sin^{2}\theta_{W}\right)\label{Eq.25}\end{equation}

They take the particular forms in particular representations, namely:

\begin{equation}
Q^{(\mathbf{3^{*}},0)}(Z_{\mu}^{1})=\frac{e}{\sin2\theta_{W}}\left(\begin{array}{ccc}
-\frac{2\sqrt{3-4\sin^{2}\theta_{W}}}{3} & 0 & 0\\
0 & \frac{\sqrt{3-4\sin^{2}\theta_{W}}}{3} & 0\\
0 & 0 & \frac{\sqrt{3-4\sin^{2}\theta_{W}}}{3}\end{array}\right)\label{Eq.26}\end{equation}

\begin{equation}
Q^{(\mathbf{3},+\frac{1}{3})}(Z_{\mu}^{1})=\frac{e}{\sin2\theta_{W}}\left(\begin{array}{ccc}
\frac{6-10\sin^{2}\theta_{W}}{3\sqrt{3-4\sin^{2}\theta_{W}}} & 0 & 0\\
0 & \frac{-3+2\sin^{2}\theta_{W}}{3\sqrt{3-4\sin^{2}\theta_{W}}} & 0\\
0 & 0 & \frac{-3+2\sin^{2}\theta_{W}}{3\sqrt{3-4\sin^{2}\theta_{W}}}\end{array}\right)\label{Eq.27}\end{equation}

\begin{equation}
Q^{(\mathbf{3},-\frac{1}{3})}(Z_{\mu}^{1})=\frac{e}{\sin2\theta_{W}}\left(\begin{array}{ccc}
\frac{2(1-\sin^{2}\theta_{W})}{\sqrt{3-4\sin^{2}\theta_{W}}} & 0 & 0\\
0 & \frac{2\sin^{2}\theta_{W}-1}{\sqrt{3-4\sin^{2}\theta_{W}}} & 0\\
0 & 0 & \frac{2\sin^{2}\theta_{W}-1}{\sqrt{3-4\sin^{2}\theta_{W}}}\end{array}\right)\label{Eq.28}\end{equation}

\begin{equation}
Q^{(\mathbf{3^{*}},0)}(Z_{\mu}^{2})=\frac{e}{\sin2\theta_{W}}\left(\begin{array}{ccc}
\frac{2}{3}\sin^{2}\theta_{W} & 0 & 0\\
0 & -1+\frac{2}{3}\sin^{2}\theta_{W} & 0\\
0 & 0 & 1-\frac{4}{3}\sin^{2}\theta_{W}\end{array}\right)\label{Eq.29}\end{equation}

\begin{equation}
Q^{(\mathbf{3},+\frac{1}{3})}(Z_{\mu}^{2})=\frac{e}{\sin2\theta_{W}}\left(\begin{array}{ccc}
-\frac{4}{3}\sin^{2}\theta_{W} & 0 & 0\\
0 & 1-\frac{4}{3}\sin^{2}\theta_{W} & 0\\
0 & 0 & -1+\frac{2}{3}\sin^{2}\theta_{W}\end{array}\right)\label{Eq.30}\end{equation}

\begin{equation}
Q^{(\mathbf{3},-\frac{1}{3})}(Z_{\mu}^{2})=\frac{e}{\sin2\theta_{W}}\left(\begin{array}{ccc}
0 & 0 & 0\\
0 & 1 & 0\\
0 & 0 & 2\sin^{2}\theta_{W}-1\end{array}\right)\label{Eq.31}\end{equation}

The last expression (Eq.31) corresponds obviously to the neutral charges
of the leptons with respect to the neutral boson from SM. The identification
stands $Z_{\mu}^{2}=Z_{\mu}^{0}$ and $Z_{\mu}^{1}=Z_{\mu}^{\prime0}$.
To complete the tableau, one computes also the charges of the singlets:
$Q^{(\mathbf{1,-1})}(Z_{\mu}^{0})=\frac{2e}{\sin2\theta_{W}}\sin^{2}\theta_{W}$,
$Q^{(\mathbf{1,-1})}(Z_{\mu}^{\prime0})=\frac{2e}{\sin2\theta_{W}\sqrt{3-4\sin^{2}\theta_{W}}}\sin^{2}\theta_{W}$,
$Q^{(\mathbf{1,-\frac{1}{3}})}(Z_{\mu}^{0})=\frac{2e{}}{3\sin2\theta_{W}}\sin^{2}\theta_{W}$,
$Q^{(\mathbf{1,-\frac{1}{3}})}(Z_{\mu}^{\prime0})=\frac{2e}{3\sin2\theta_{W}\sqrt{3-4\sin^{2}\theta_{W}}}\sin^{2}\theta_{W}$,
$Q^{(\mathbf{1,+\frac{2}{3}})}(Z_{\mu}^{0})=-\frac{4e{}}{3\sin2\theta_{W}}\sin^{2}\theta_{W}$,
$Q^{(\mathbf{1,+\frac{2}{3}})}(Z_{\mu}^{\prime0})=-\frac{4e}{3\sin2\theta_{W}\sqrt{3-4\sin^{2}\theta_{W}}}\sin^{2}\theta_{W}$. 

Now all the currents in the 3-3-1 model with right-handed neutrinos
are exactly computed and their couplings are summarized in the Table
1. We mention that these results at the tree level need no other adjustments,
since the mixing angle $\phi$ (that appears for instance in \cite{key-5})
was in our method aleready taken into consideration when the neutral
bosons were rotated (through the matrix $\omega$) to their mass eigenstates.
Anyhow, Ref. \cite{key-5} obtaines the same couplings in the limit
$\phi\rightarrow0$ (which is quite plausibleand natural, since the
experimental values invoked therein for $\phi$ are in the vecinity
of zero). 

\begin{table}

\caption{Coupling coefficients of the neutral currents in a 3-3-1 model with
right-handed neutrinos}

\begin{tabular}{cccc}
\hline 
Particle\textbackslash{}Coupling($e/\sin2\theta_{W}$)&
$Z\rightarrow\bar{f}f$&
&
$Z^{\prime}\rightarrow\bar{f}f$\tabularnewline
\hline
\hline 
$e_{L},\mu_{L},\tau_{L}$&
$2\sin^{2}\theta_{W}-1$&
&
$\frac{2\sin^{2}\theta_{W}-1}{\sqrt{3-4\sin^{2}\theta_{W}}}$\tabularnewline
&
&
&
\tabularnewline
$\nu_{eL},\nu_{\mu L},\nu_{\tau L}$&
$1$&
&
$\frac{2\sin^{2}\theta_{W}-1}{\sqrt{3-4\sin^{2}\theta_{W}}}$\tabularnewline
&
&
&
\tabularnewline
$e_{R},\mu_{R},\tau_{R}$&
$2\sin^{2}\theta_{W}$&
&
$\frac{2\sin^{2}\theta_{W}}{\sqrt{3-4\sin^{2}\theta_{W}}}$\tabularnewline
&
&
&
\tabularnewline
$\nu_{eR},\nu_{\mu R},\nu_{\tau R}$&
$0$&
&
$\frac{2(1-\sin^{2}\theta_{W})}{\sqrt{3-4\sin^{2}\theta_{W}}}$\tabularnewline
&
&
&
\tabularnewline
$u_{L},c_{L}$&
$1-\frac{4}{3}\sin^{2}\theta_{W}$&
&
$\frac{1}{3}\sqrt{3-4\sin^{2}\theta_{W}}$\tabularnewline
&
&
&
\tabularnewline
$d_{L},s_{L}$&
$-1+\frac{2}{3}\sin^{2}\theta_{W}$&
&
$\frac{1}{3}\sqrt{3-4\sin^{2}\theta_{W}}$\tabularnewline
&
&
&
\tabularnewline
$t_{L}$&
$1-\frac{4}{3}\sin^{2}\theta_{W}$&
&
$\frac{-1+\frac{2}{3}\sin^{2}\theta_{W}}{\sqrt{3-4\sin^{2}\theta_{W}}}$\tabularnewline
&
&
&
\tabularnewline
$b_{L}$&
$-1+\frac{2}{3}\sin^{2}\theta_{W}$&
&
$\frac{-1+\frac{2}{3}\sin^{2}\theta_{W}}{\sqrt{3-4\sin^{2}\theta_{W}}}$\tabularnewline
&
&
&
\tabularnewline
$D_{L},S_{L}$&
$\frac{2}{3}\sin^{2}\theta_{W}$&
&
$-\frac{2}{3}\sqrt{3-4\sin^{2}\theta_{W}}$\tabularnewline
&
&
&
\tabularnewline
$T_{L}$&
$-\frac{4}{3}\sin^{2}\theta_{W}$&
&
$\frac{6-10\sin^{2}\theta_{W}}{3\sqrt{3-4\sin^{2}\theta_{W}}}$\tabularnewline
&
&
&
\tabularnewline
$u_{R},c_{R},t_{Rr},T_{R}$&
$-\frac{4}{3}\sin^{2}\theta_{W}$&
&
$\frac{-4\sin^{2}\theta_{W}}{3\sqrt{3-4\sin^{2}\theta_{W}}}$\tabularnewline
&
&
&
\tabularnewline
$d_{R},s_{R},b_{R},D_{R},S_{R}$&
$\frac{2}{3}\sin^{2}\theta_{W}$&
&
$\frac{2\sin^{2}\theta_{W}}{3\sqrt{3-4\sin^{2}\theta_{W}}}$\tabularnewline
\hline
\hline 
&
&
&
\tabularnewline
\end{tabular}
\end{table}

\section{Concluding Remarks}

Based on the algebraical approach designed to exactly solve gauge
models with high symmetries, the currents (both the electric and the
neutral ones) of the 3-3-1 model with right-handed neutrinos are explicitly
obtained here. Moreover, their exact expressions are obtained after
the rotation $\phi$ corresponding to the mixing between neutral bosons
$Z$ and $Z^{\prime}$ has been performed (not before this goal is
achieved, as in the most papers in the literature). Therefore, no
adjustment to the resulting exressions at tree level is required anymore. 

As expected, all the particles coming from the SM recover the couplings
(to the electromagnetic field and the Weinberg neutral boson respectively)
at their values established therein. At the same time, if those particles
belong to the same triplet of the model or to equivalent representations
with respect to the gauge group, they exhibit indistinguishable couplings
to the new neutral boson of the 3-3-1 model. The exotic quarks have
pure vector couplings to the Weinberg neutral boson, while the right-handed
neutrinos remain uncoupled to it. 

The flavor changing neutral currents (FCNC) and their suppresion (via
GIM mechanism) have been already discussed in the literature and our
results do not changes those conclusions, as long as we do not consider
here explicitly the Higgs fields which supplies masses to the fermions
involved in the model. It is well established that GIM mechanism occurs
only for the fermions with equivalent representations with respect
to the gauge group that, in addition, aquire masses from a unique
source. Assuming that our SSB mechanism (with one step $331\rightarrow31$)
formaly differs from the traditional one (with two steps $331\rightarrow321\rightarrow31$)
and the Yukawa terms could arise like a tensor product of various
Higgs triplets, the analysis of FCNC deserves a special atention and
it remains to be performed elsewhere. 

In conclusion, we proved in this paper that the general method of
exactly solving gauge models with high symmetries gives valuable results
regarding the charged and neutral currents in 3-3-1 models with right-handed
neutrinos in good accord with other methods. Adding to this (from
our previous papers) three arguments - (1) the method itself can enforce
the fermion family representations just by choosing a set of parameters;
(2) it can determine the mass spectrum of all the particles in the
model simply by tuning a unique parameter (due to a special SSB mechanism);
(3) it offers a lot of ways to embed the neutrino phenomenology in
the model - we consider this method a powerful theoretical tool in
investigating models with even larger gauge groups such as $SU(3)_{c}\otimes SU(4)_{L}\otimes U(1)_{X}$.

\end{document}